\documentclass[onecolumn]{emulateapj}

\newcommand{\be}{\begin{equation}}
\newcommand{\ee}{\end{equation}}
\newcommand{\bea}{\begin{eqnarray}}
\newcommand{\eea}{\end{eqnarray}}
\newcommand{\rmscr}[1]{{\hbox{\scriptsize \rm{#1}}}}
\newcommand{\rmmat}[1]{{\hbox{\rm{#1}}}}

\def\etal{{\em et al.\ }}

\slugcomment{Submitted to ApJ}

\shorttitle{Images of SgrA*}

\begin{document}
\bibliographystyle{apj}

\title{Images of the radiatively inefficient accretion flow surrounding
a Kerr black hole: application in Sgr A*}

\author{Ye-Fei Yuan\altaffilmark{1}, Xinwu Cao\altaffilmark{2}, Lei Huang\altaffilmark{1,2,3}and Zhi-Qiang Shen
\altaffilmark{2}}

\altaffiltext{1}{Key Laboratory for Research in Galaxies and
Cosmology, University of Sciences and Technology of China,
Chinese Academy of Sciences, Hefei 230026, China;
yfyuan,mlhuang@ustc.edu.cn}

\altaffiltext{2}{Key Laboratory for Research in Galaxies and
Cosmology, Shanghai Astronomical Observatory, Chinese Academy of
Sciences, Shanghai 200030, China; cxw,zshen@shao.ac.cn}

\altaffiltext{3}{Academia Sinica, Institute of Astronomy and
Astrophysics, Taipei 106, Taiwan}

\begin{abstract}
In fully general relativity, we calculate the images of the
radiatively inefficient accretion flow (RIAF) surrounding a Kerr
black hole with arbitrary spins, inclination angles, and
observational wavelengths. For the same initial conditions, such as
the fixed accretion rate, it is found that the intrinsic size and
radiation intensity of the images become larger, but the images
become more compact in the inner region, while the size of the black
hole shadow decreases with the increase of the black hole spin. With
the increase of the inclination angles, the shapes of the black
hole shadows change and become smaller, even disappear at all due to
the obscuration by the thick disks. For median inclination angles,
the radial velocity observed at infinity is larger because of both
the rotation and radial motion of the fluid in the disk, which
results in the luminous part of the images is much brighter. For
larger inclination angles, such as the disk is edge on, the emission
becomes dimmer at longer observational wavelengths (such as at 7.0mm
and 3.5mm wavelengths), or brighter at shorter observational
wavelengths (such as at 1.3 mm wavelength) than that of the face on
case, except for the high spin and high inclination images. 
These complex behaviors are due to the combination of the
Lorentz boosting effect and
the radiative absorption in the disk. We hope our results are
helpful to determine the spin parameter of the black hole in low
luminosity sources, such as the Galactic center. A primary
analysis by 
comparison with the observed sizes of Sgr A* at millimeters strongly
suggests that the disk around the central black hole at Sgr A* 
is  highly inclined or the central black hole is rotating fast.

\end{abstract}


\section{Introduction}
More and more observations indicate that supermassive black holes
reside in most local galaxies, and the co-evolution of galaxies,
quasars and supermassive black holes was realized in last decade
\citep[e.g.,][]{2000ApJ...539L...9F,2000ApJ...539L..13G,2008ApJS..175..356H}.
From the stellar dynamics, it is not surprising to find that the
compact radio source Sagittarius (Sgr) A* at the center of our
Galaxy is associated with a $\sim 4 \times 10^6 M_{\sun}$ massive
dark object
\citep{2003ApJ...596.1015S,2005ApJ...620..744G,2008ApJ...689.1044G}.
A natural candidate of the massive dark object is massive black
hole, as emphasized by \citet{1995ARA&A..33..581K}, they argued
convincingly that this object is a black hole, and the other
alternatives to a black hole should be ruled out. Some of these
alternatives have already been excluded. They include a cluster of
dark objects, such as the stellar remnants or brown dwarfs
\citep{1998ApJ...494L.181M}, a binary made of point masses, self
gravitationally supported Fermi ball
\citep{2006MNRAS.367L..32M,1998ApJ...500..591T,1998ApJ...509L.105M,2004ApJ...606.1112Y}.
Another alternative which is still waiting for being excluded is
self gravitationally supported boson ball
\citep{2000PhRvD..62j4012T,2004ApJ...606.1112Y}. However, the
problem of this model is that the mass of bosons is arbitrarily
assumed to make a massive boson ball, and the energy dissipation
between bosons may lead to the collapse of the ball
\citep{1998ApJ...494L.181M}.

Recent very-long baseline interferometry ({\it VLBI}) observation
at the wavelength of 3.5mm shows its intrinsic size is
0.126$\pm$0.017 mas, about $\sim 1AU$, and the resulting lower
limit of its mass density is $6.5 \times 10^{21} M_{\sun} pc^{-3}$
\citep{2005Natur.438...62S}, and the {\it VLBI} observation at
1.3mm gives the intrinsic size $0.037^{+0.016}_{-0.010}$ mas, and
the resulting lower limit of its mass density is $9.3 \times
10^{22} M_{\sun} pc^{-3}$ \citep{2008Natur.455...78D}, which
strongly supports that Sgr A* is a supermassive black hole. The
millimeter {\it VLBI} observation has been used to estimate the
parameters of SgrA* accretion flow \citep{2008arXiv0809.4490B}. In
the near future, higher resolution observations of Sgr A* at
infrared \citep{2005AN....326..568P} and sub-millimeter
wavelengths will be 
available \citep{2004GCNew..18....6D,2004PThPS.155..186M,2008arXiv0809.4489F},
which can provide a direct test of the physical processes under
the strong gravity, such as the dynamics of the accreted gas,
complex trajectories of photons, and so on. Therefore, it is
necessary to investigate the images of the flux and polarization
of realistic accretion flows
\citep{2004ApJ...606..894Y,2006ApJ...636L.109B,2007MNRAS.379..833H,2008ApJ...676L.119H},
and the images and light curves associated with a hot spot in the
accretion flow near the black hole horizon
\citep{2005MNRAS.363..353B,2006MNRAS.367..905B}, the hot spot is
proposed to explain the observations of near-infrared and X-ray
flaring of Sgr A*.

It is generally believed that Sgr A* was a low luminosity AGN
whose activity was switched off due to insufficient gas supply
\citep{2002luml.conf..405N}. Multi-wavelength observations put
strict constraints on theoretical accretion models of the massive
black hole at the Galactic center. Among the theoretical models,
the advection-dominated accretion flow (ADAF) model, also called
the radiatively inefficient accretion flow (RIAF) model is claimed
to be very successful in modeling the multi-wavelength spectrum
from radio to X-ray
\citep{1995Natur.374..623N,1998ApJ...492..554N,2003ApJ...598..301Y,2004ApJ...606..894Y}.
Based on the RIAF model, the images of Sgr A* at millimeter have
been calculated,  the size of the images is compared with
observations to put an independent constraint on the RIAF model
\citep{2006ApJ...642L..45Y,2007MNRAS.379..833H}, and the shape of
the images, especially the location of the image centroid of Sgr
A* can be used to determine the black hole spin
\citep{2006ApJ...636L.109B}.

There are several shortcomings in the previous calculations: first,
although the treatment of the photon trajectories is within the
framework of fully general relativity by using a ray tracing method,
the accretion model is Newtonian, which means the global structure
of the accretion flow is based on the Newtonian dynamics.  The
general relativity effect for the accretion flow is mimicked with
Paczyn´ski \& Wiita potential. Therefore, the radial velocity of
the accretion flow very close to the black hole is larger than the
speed of light and thus is unphysical. Second, the massive black
hole is assumed to be a non-rotating (Schwarzschild) hole in
\citet{2004ApJ...606..894Y}.

In this paper, following \citet{2000ApJ...534..734M}, the global
structure of the relativistic ADAF around a Kerr black hole is
derived to calculate the images of Sgr A* surrounding a Kerr black
hole with an arbitrary black hole spin and viewing angle at several
millimeter wavebands. Our treatment is within the framework of fully
general relativity. A brief introduction to relativistic ADAF model
is given in \S2. The ray tracing method and the radiation transfer
in curved space-time are discussed in \S3. In \S4, we summarize the
key steps for calculating the images of ADAF of Sgr A*. The results
and discussions are presented in \S5 and \S6, respectively.

\section{General Relativistic RIAF Model}
In the Boyer-Lindquist coordinates $(t,r,\theta,\phi)$,
the Kerr metric is
\be
ds^2=-e^{2\nu}dt^2+e^{2\psi}(d\phi-\omega dt)^2 +e^{2\mu_1}dr^2+e^{2\mu_2}d\theta^2,
\ee
where
\begin{equation}
e^{2\nu}=\frac{\Sigma \Delta}{A}, \,\,\,
e^{2\psi}=\frac{A \sin ^2 \theta}{\Sigma}, \,\,\,
e^{2\mu _1}=\frac{\Sigma}{\Delta}, \,\,\,
e^{2\mu_2}=\Sigma, \,\,\,
\omega=\frac{2ar}{A},  \,\,\,
\end{equation}
\be
\Delta=r^2-2r+a^2,  \,\,\,
\Sigma=r^2+a^2\cos ^2\theta, \,\,\,
A=(r^2+a^2)^2-a^2\Delta\sin ^2\theta .
\ee
Here, the notation in the seminal paper
\citep{1972ApJ...178..347B} has been taken, and the geometrical
units $G=c=1$ are used throughout this paper.
The accretion flow has both azimuthal and radial motion in the
accretion flow.
The velocity field of the fluid is described by $\Omega$ and $V$, where
$\Omega$ is the azimuthal angular velocity of the fluid and
$V$ is the physical velocity of the fluid with respect to the co-rotating
reference frame which is carried by the observers with the four velocity
$u^{\mu}=u^t(1,0,0,\Omega)$
\citep{1997ApJ...479..179A,1997MNRAS.286..681P,1998ApJ...498..313G,2000ApJ...534..734M}.
The four velocity of the fluid is (see the
Appendix for the details)
\be
u^{\mu}=(\gamma_r \gamma_{\phi} e^{-\nu},
         \gamma_r \beta_r e^{-\mu_1},
         0,
         u^t \Omega).
\ee

In order to solve the global solution of the RIAF in Kerr metric, we
follow \citet{2000ApJ...534..734M} to formulate the fully
relativistic hydrodynamical equations for RIAF. For simplicity, they
assume the height of the accretion flow $H\leq r$, therefore the
Boyer-Lindquist coordinates $(t, r, \theta, \phi)$ are expanded
around the equatorial plane up to $(z/r)^0 $ terms to obtain a
cylindrical coordinates $(t,\tilde{r},\phi, z)$. Another
approximation is generally taken in previous investigations: the
vertical integration. Therefore, all the physical quantities are
taken their values at the equatorial plane. We employ the approach
suggested by \citet{2000ApJ...534..734M} to calculate the global
structure of an accretion flow surrounding a massive black hole in
the general relativistic frame, which allows us to calculate the
structure of an accretion flow surrounding either a spinning or a
non-spinning black hole. All the radiation processes are included in
the calculations of the global accretion flow structure
\citep*[see][for the details and the references
therein]{2000ApJ...534..734M}. Integrating these equations from the
outer boundary of the flow at $r=r_{\rm out}$ inwards the black
hole, we can obtain the global structure of the accretion flow
passing the sonic point smoothly to the black hole horizon. Here, we
allow the accretion rate of the accretion flow to vary with radius
due to the putative winds \citep*[see][for the
details]{1999MNRAS.303L...1B}, i.e., we assume the accretion rate
$\dot{M}(r)=\dot{M}_0(r/ r_{\rm out})^s$, where $\dot{M}_0$ is the
accretion rate at the outer radius of the ADAF
\citep{2003ApJ...598..301Y}.

\begin{figure}
\includegraphics[angle=-90,scale=.7]{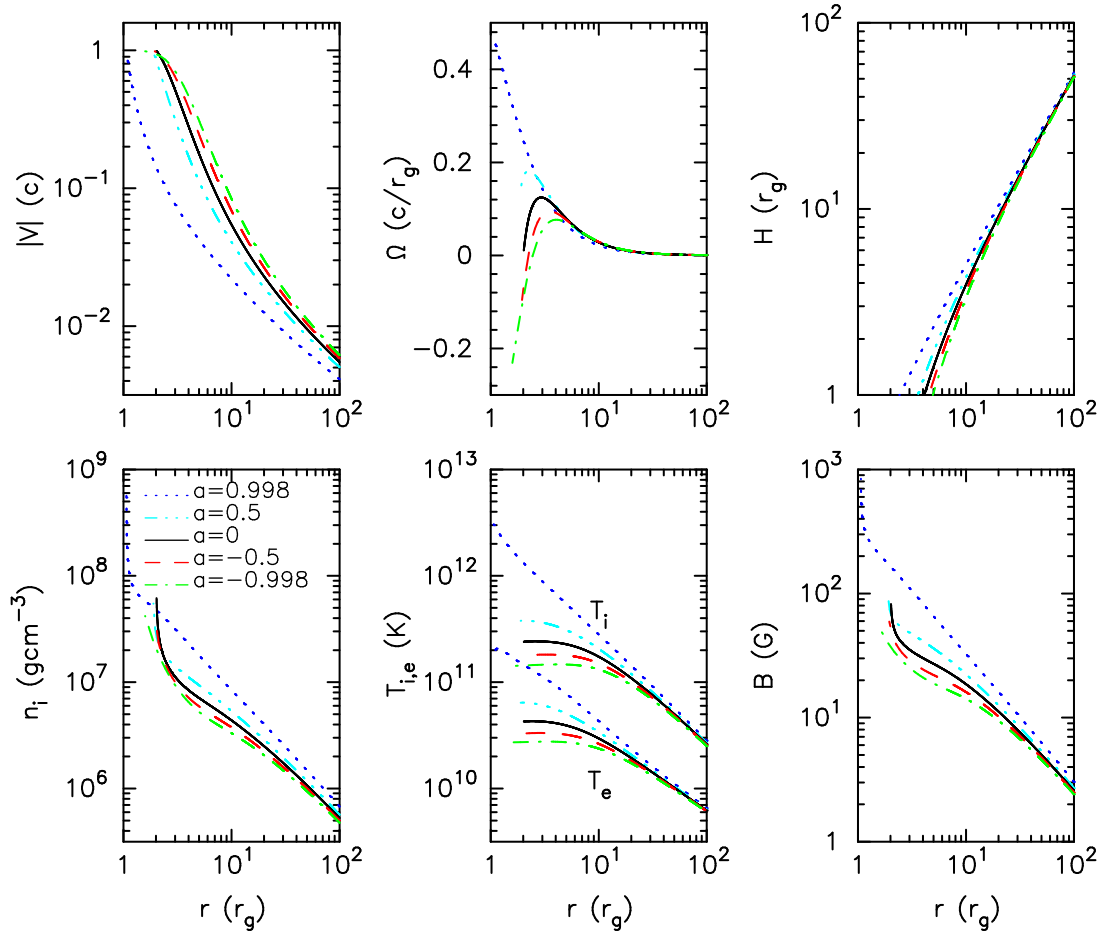}
\caption{Global solution of GR-ADAF.
The solid lines represent the results for $a=0$,
the dotted lines the results for $a=0.998$,
the dotted-dotted-dashed lines the results for $a=0.5$,
the dashed lines the results for $a=-0.5$, and
the dotted-dashed lines the results for $a=-0.998$.
Except the upper left panel, the lines above the solid ones
represent the results for prograde disks ($a>0$), while
the lines below the results for retrograde disks ($a<0$).
{\it Upper left:} Radial velocity $V$ as a function of radii.
The lines below the solid ones are for $a>0$,
while the lines above are for $a<0$.
{\it Upper middle:} Coordinate angular velocity $\Omega$.
{\it Upper right:} The half height of the disk $H$.
{\it Lower left:} The number density of ions versus radii.
{\it Lower middle:} The temperature of ions and electrons.
{\it Lower right:} The magnetic field $B$ which is deduced from the
global solution of ADAF.
\label{fig_adaf}}
\end{figure}
The global solution of general relativistic RIAF
(GR-RIAF hereafter) has been obtained in the previous
works, and the effects of black hole rotation have been discussed
in the great details \citep{1998ApJ...498..313G,1998ApJ...504..419P,2000ApJ...534..734M}.
In order to help understand our numerical results, the global
solution of ADAF is shown in Fig.\ref{fig_adaf}.
The basic parameters are taken as follows: the black hole mass of
Sgr A* $M_{\rm bh}=4\times 10^6 M_\odot$, the wind index $s=0.25$,
the fraction of the dissipated power directly heating the electrons
$\delta=0.3$, the accretion rate at the outer radius
$\dot{M}_0=0.7\times10^{-6} M_{\odot} {\rm yr}^{-1}$, the ratio of
the gas pressure to the total pressure $\beta=0.95$. These
parameters are the same as the RIAF parameters in
\citet{2006ApJ...642L..45Y}'s work, except the slightly difference
in accretion rate (see \S \ref{sec:results} for the discussion). It
should be emphasized that the ion and electron temperatures increase
significantly within $10^2 R_g$ with the increase of the black hole
spin parameter from $a=0$ to $a=0.998$, and decrease, with less
difference, with its decrease from $a=0$ to $a=-0.998$. Similar
changes appear in electron number density and magnetic field
strength. Because the radio emission is mainly from the synchrotron
radiation which depends sensitively on electron temperature, the
effects of the black hole spin on the radio emission is mainly
through its effects on the electron temperature.

\section{Relativistic Radiation Transfer of Photons from GR-RIAF}
In this paper, the images of Sgr A*  modeled as a GR-RIAF
surrounding a massive black hole are calculated. In our
calculations, we do not consider the Compton effect of photons at
radio wavelength, because the relativistic electrons will boost
the low energy photons to higher energy which is beyond our interest. As a
bunch of photons travel along their geodesics, some of them will
be absorbed due to synchrotron self-absorption in the accretion
flow, while some of new photons will be added due to the emission
of the flow. To solve the radiation transfer of radio photons, we
should first determine their trajectories using the ray tracing
method.

\subsection{Ray Tracing Method}
In the Kerr space-time, all of the four constants for the motion of
a test particle have been found by \citet{1968PhRv..174.1559C}, they
are the energy at infinity \be E=-p_t, \ee the effective angular
momentum, \be L_z=p_{\phi}, \ee the Carter constant, \be
Q=p^2_{\theta}-a^2 E^2\cos^2\theta + L_z^2 \cot^2 \theta, \ee and
the Hamiltonian constant, \be
H=\frac{1}{2}\epsilon=\frac{1}{2}p^{\mu}p_{\mu}, \ee where
$\epsilon=0$ for massless particles such as photons, or $\epsilon=1$
for massive particles. Therefore, given three constants of motion,
i.e. $E, L_z, Q$, the trajectory of a photon can be determined. As
the trajectory of a photon is independent on its energy, it can be
completely determined by the two dimensionless parameters
$\lambda=L_z/E$ and $q=Q^{1/2}/E$. Furthermore,
\citet{1973ApJ...183..237C} pointed out that the two parameters
$\lambda$ and $Q$ are related to the two impact parameters $\alpha$
and $\beta$ that describe the apparent position of the image on the
sky as seen by an observer at infinity who receives that light ray,
\be \alpha=-\left(\frac{rp^{(\phi)}}{p^{(t)}}\right)_{r \rightarrow
\infty}
     =-\frac{\lambda}{\sin \theta_{\rmscr{obs}}}, \label{eq:alpha}
\ee
\be
\beta=\left(\frac{rp^{(\theta)}}{p^{(t)}}\right)_{r \rightarrow \infty}
    =(q+a^2\cos^2\theta_{\rmscr{obs}}-\lambda^2\cot^2\theta_{\rmscr{obs}})^{1/2}
    =p_{\rmscr{obs}}, \label{eq:beta}
\ee where $\theta_{\rmscr{obs}}$ is the angular coordinate of the
observer at infinity and $p^{(a)}$ is the four momentum in the
locally non-rotating frame (see the Appendix). The direction of
$\beta$ points along the projected direction of the spin axis of
black hole. Due to the azimuthal symmetry, the photon trajectory
originating from a position $(r, \theta)$ to the observer at
infinity whose angular coordinate is $\theta_{\rmscr{obs}}$ can be
solved by the following integral equation: \be \pm \int_{r}^{\infty}
\frac{dr}{\sqrt{R(r;\lambda,q)}} =\pm
\int_{\theta}^{\theta_{\rmscr{obs}}}
\frac{d\theta}{\sqrt{\Theta(\theta;\lambda,q)}}=P, \label{eq:traj}
\ee where \bea
R(r;\lambda,q)&=&r^4+(a^2-\lambda^2-q)r^2+2[q+(\lambda-a)^2]r-a^2q, \\
\Theta(\theta;\lambda,q)&=&(q+a^2\cos^2\theta-\lambda^2\cot^2\theta).
\eea
The signs in the above integrals must be the same as those of
$dr$ and $d\theta$ to guarantee the integral $P$ is always positive and
increasing along a photon's trajectory. Therefore, the signs change at
the turning points in $r$ and $\theta$, where $R(r;\lambda,q)=0$ or
$\Theta(\theta;\lambda,q)=0$ respectively.
It is important to realize
that both $R(r)$ and $\Theta(\theta)$ are quartic
polynomials of $r$ and $\theta$ respectively, therefore
both integrals in Eq.\,(\ref{eq:traj}) can be integrated in terms of
elliptic functions
\citep{1983mtbh.book.....C,1994ApJ...421...46R,1998NewA....3..647C}.

Owing to above observations, a photon trajectory can be determined
by using a ray tracing method: first, given $\alpha$ and $\beta$
which determine the position of the image in the photo-graphical
plate of an observer with $\theta_{\rmscr{obs}}$, the two constants
of motion are obtained, according to
Eqs.\,(\ref{eq:alpha}-\ref{eq:beta}). Second, given the integral
$P$, which can be understood as the affine parameter along the
photon's trajectory, the position of the photon can be determined
using the expression of Eq.(\ref{eq:traj}) as elliptic functions:
\bea
r&=&r(P;\lambda,q), \\
\theta&=&\theta(P;\lambda,q).
\eea

\subsection{Relativistic Radiation Transfer}
Given the two constants of motion of a photon, its momentum has
components, \be k_{\mu}=g_{\mu
\nu}\frac{dx^{\nu}}{d\sigma}=(k_t,k_r,k_{\theta},k_{\phi})=
\left(-1,\pm \frac{\sqrt{R}}{\Delta}, \pm \sqrt{\Theta},\lambda
\right) \ee where $\sigma$ is the affine parameter of photon's
trajectory, and $E$ is absorbed into $\sigma$. The signs in the
above equation are the same as those of $dr$ and $d\theta$.
Suppose the photon frequency observed by the observer at infinity
is $\nu_{\rmscr{obs}}$, then the emitting frequency at the locally
rest frame of the fluid (LRF) is \be
\nu_{\rmscr{em}}=-(k_{\mu}u^{\mu}) \nu_{\rmscr{obs}}, \ee where
$u^{\mu}$ is the four velocity of the fluid. In the LRF, the
radiation transfer equation is \citep{1979rpa..book.....R} \be
\frac{dI(\nu_{\rmscr{em}})}{dL}=-\kappa(\nu_{\rmscr{em}})
I(\nu_{\rmscr{em}}) + \eta(\nu_{\rmscr{em}}), \ee where $I$ is the
radiation intensity of photons and $dL$ is the element of physical
distance of photon traveling in LRF. As ${\cal I}\equiv I/\nu^3$
is a Lorentz invariant, the above equation can be re-written as
follows, \be \frac{d{\cal
I(\nu_{\rmscr{em}})}}{dL}=-\kappa(\nu_{\rmscr{em}}){\cal
I(\nu_{\rmscr{em}})}
+\frac{\eta(\nu_{\rmscr{em}})}{\nu_{\rmscr{em}}^3}. \ee Since the
speed of light is $c=1$, we have \be
\frac{dL}{d\sigma}=\frac{dx^{(0)}}{d\sigma}=\frac{dx^{\mu}}{d\sigma}e^{(0)}_{\mu}
({\rm LRF}) =-k_{\mu}u^{\mu}. \ee Therefore, the radiation
transfer equation in curved space-time reads
\citep{2004A&A...424..733F},

\bea
\frac{d{\cal I(\nu_{\rmscr{em}})}}{d\sigma}&=&-k_{\mu}u^{\mu}
[-\kappa(\nu_{\rmscr{em}}) {\cal I(\nu_{\rmscr{em}})}+\eta(\nu_{\rmscr{em}})/\nu_{\rmscr{em}}^3]  \\
&=& [-\nu_{\rmscr{em}} \kappa(\nu_{\rmscr{em}}) {\cal I(\nu_{\rmscr{em}})}+\eta(\nu_{\rmscr{em}})/\nu_{\rmscr{em}}^2]/\nu_{\rmscr{obs}}. \label{eq:transfer}
\eea
It is well known that ${\cal I}, \nu \kappa(\nu)$, and $\eta(\nu)/\nu^2$
are Lorentz invariants \citep{1979rpa..book.....R},
therefore, the above radiation transfer equation
is written in the Lorentz invariant form as we expected.

Since
\bea
\Sigma \frac{dr}{d\sigma} &=& \pm \sqrt{R}, \\
\frac{dr}{dP} &=& \pm \sqrt{R}, \eea therefore, \be d\sigma = -
\Sigma dP. \ee Introducing $\widetilde{\kappa}=\nu_{\rmscr{em}}
\kappa(\nu_{\rmscr{em}}) /\nu_{\rmscr{obs}}$ and
$\widetilde{\eta}=\eta(\nu_{\rmscr{em}})/ (\nu_{\rmscr{em}}^2
\nu_{\rmscr{obs}})$, and the formal solution to the transfer
equation reads,
\be
{\cal I}(\infty)=\int_0^{P_{\rmscr{max}}} \exp
\left(-\int_{P^{'}}^{P_{\rmscr{max}}} \widetilde{\kappa}(P^{''})\Sigma dP^{''}\right)
\widetilde{\eta}(P^{'}) \Sigma dP^{'} , \label{eq:flux}
\ee
where $P_{\rmscr{max}}$ is the maximal trajectory integral tracing from the infinity
to the innermost emitting position.

\section{Images of GR-RIAF at Millimeter Wavelengths }
After solving the global structure of the RIAF around a
Schwarzschild or Kerr black hole, the physical quantities at the
equatorial plane have been obtained, such as: \be
n_e(\tilde{r},z=0),n_i(\tilde{r},z=0),V_r(\tilde{r},z=0),\Omega(\tilde{r},z=0).
\ee In the following calculations, the isothermal approximation is
taken in the vertical direction, therefore, the density distribution
in the vertical direction is: \bea
n_e(\tilde{r},z)&=&n_e(\tilde{r},z=0)\exp(-z^2/2H^2), \nonumber \\
n_i(\tilde{r},z)&=&n_i(\tilde{r},z=0)\exp(-z^2/2H^2), \nonumber
\eea
and
\bea
V_r(\tilde{r},z)&=&V_r(\tilde{r},z=0), \nonumber \\
\Omega(\tilde{r},z)&=&\Omega(\tilde{r},z=0).\nonumber
\eea

At any position along the photon trajectory, such as $(r(P),\theta(P))$,
we can determine $u^{\mu}(P)$ and $k_{\mu}(P)$, so we get
$\nu_{\rmscr{em}}(P)=(-u^{\mu}(P)k_{\mu}(P))\nu_{\rmscr{obs}}$.
The magnitude of magnetic filed $B(\tilde{r},z)$ is determined by
the local physical conditions:
\begin{eqnarray}
P_{\rm gas}(\tilde{r},z)&=&n_i(\tilde{r},z)k_{\rm B}T_i+n_e(\tilde{r},z)k_{\rm B}T_e, \nonumber \\
P_{\rm mag}(\tilde{r},z)&=&P_{\rm gas}(\tilde{r},z)(1-\beta)/\beta, \nonumber \\
B(\tilde{r},z)&=&\sqrt{8\pi P_{\rm mag}} \nonumber
\end{eqnarray}

The emissivity and absorption coefficients at each position can be
calculated, considering the bremsstrahlung and synchrotron radiation
mechanism. The Compton scattering effect is ignored, because it
generally produces photons at higher wavelength band which are not
interested in this work. Finally, the radiation intensity received
by the observer at infinity can be calculated according to
Eq.(\ref{eq:flux}).

\section{Numerical Results} \label{sec:results}

In the work of \citet{2006ApJ...642L..45Y}, the size of Sgr A* was
calculated in the RIAF model of \citet{2003ApJ...598..301Y}. In
this RIAF model, the accretion rate is assumed to be
$\dot{M}(r)=\dot{M}_0(r/{\rm r_{\rm out}})^s$ due to the presence of
winds, where $\dot{M}_0$ is the accretion rate at the outer radius
of the RIAF \citep{2003ApJ...598..301Y}. The black hole mass of Sgr
A* $M_{\rm bh}=4\times 10^6 M_\odot$ is adopted in their work, which
is different from $M_{\rm bh}=2.5\times 10^6 M_\odot$ adopted in
\citet{2003ApJ...598..301Y}. Thus, they tuned the parameters of the
RIAF in order to reproduce the observed spectrum in different
wavebands, which leads to $s=0.25$, the fraction of the dissipated
power directly heating the electrons $\delta=0.3$. Their
calculations of the RIAF structure were carried out by adopting the
Pseudo-Newtonian potential. Compared with the results in fully
general relativistic frame in the case of a Schwarzschild black hole
(i.e., $a=0$), their global solution of the RIAF deviates only slightly
in region out of $\sim 20 R_g$, but to a certain extent in the inner
region within $\sim 20 R_g$. In this work, we simply adopt the same
RIAF parameters as in \citet{2006ApJ...642L..45Y} in all our
calculations independent of black hole spin parameter $a$, except
that a slightly lower accretion rate ($\dot{M}=0.7\times10^{-6}
M_{\odot} {\rm yr}^{-1}$, 70 percent of that adopted by
\citet{2006ApJ...642L..45Y}) at the outer boundary of the RIAF is
adopted in our calculations, which can compensate the discrepancy
between our GR-RIAF solution and the RIAF solution of
\citet{2006ApJ...642L..45Y} in the inner region of the flow near the
black hole with $a=0$. We find that our global GR-RIAF solution can
reproduce the broad spectrum from radio to X-ray quite well for a
non-rotating black hole while viewed face-on.
We treat this as our fiducial model.
In principle, we should
have to tune the RIAF parameters to fit the observed spectrum,
the size of the scattered images, and even the polarization
properties, whenever the black hole spin parameter $a$ and the
inclination angle $\theta_{\rmscr{obs}}$ are specified
as done in \citet{2008arXiv0809.4490B},
but this is very cumbersome and beyond the scope of this
work, which should be carried out in our later work.
To focus on the influence of black hole spin on the
general properties of the images at
millimeter wavelengths, as a first step,
we will not change any RIAF parameters in our
calculations for the GR-RIAF structure, except for different values
of black hole spin parameters. Because the non-thermal electrons
are artificially added to explain the observed long wavelength
($\gtrsim 7$mm) radio spectrum, we consider the
radio emission is contributed only from the thermal electrons in the
following calculations.

\begin{figure}
\includegraphics[scale=0.99]{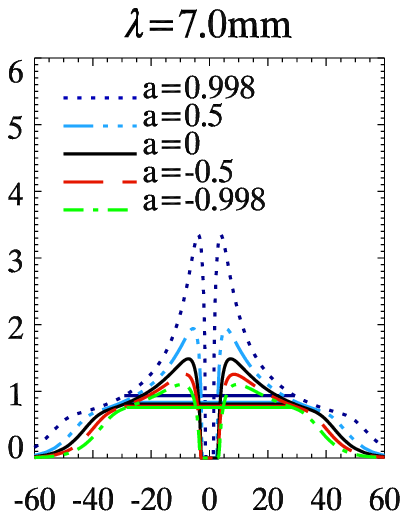}
\includegraphics[scale=0.99]{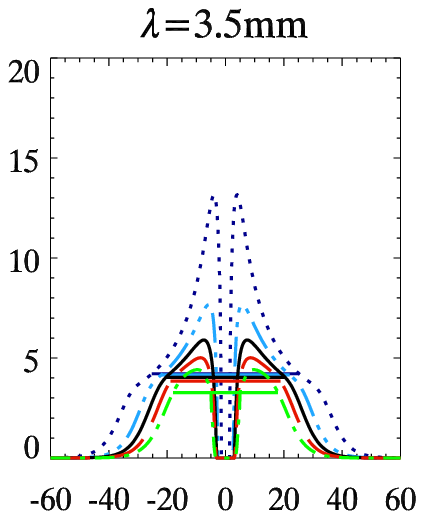}
\includegraphics[scale=0.99]{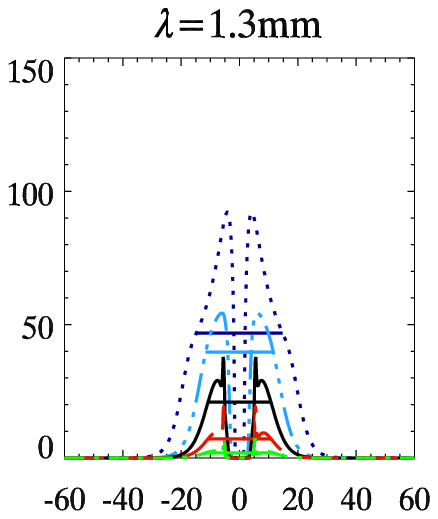}
\caption{Intensity of the ADAF around Kerr black holes with
different spins at the wavelengths of $\lambda=7.0mm, 3.5mm, 1.3mm$
(columns from left to right)
as a function of $\alpha$ (or $\beta$) when $\theta_{\rmscr{obs}}=0^0$.
Results for different spin parameter $a$ are shown in line styles the same
as those in Fig.\ref{fig_adaf}. The horizontal lines show the intrinsic
size of the disk which is defined according to Eq.(28)-(29).
\label{fig_0}}
\end{figure}
\begin{figure}
\includegraphics[scale=.42]{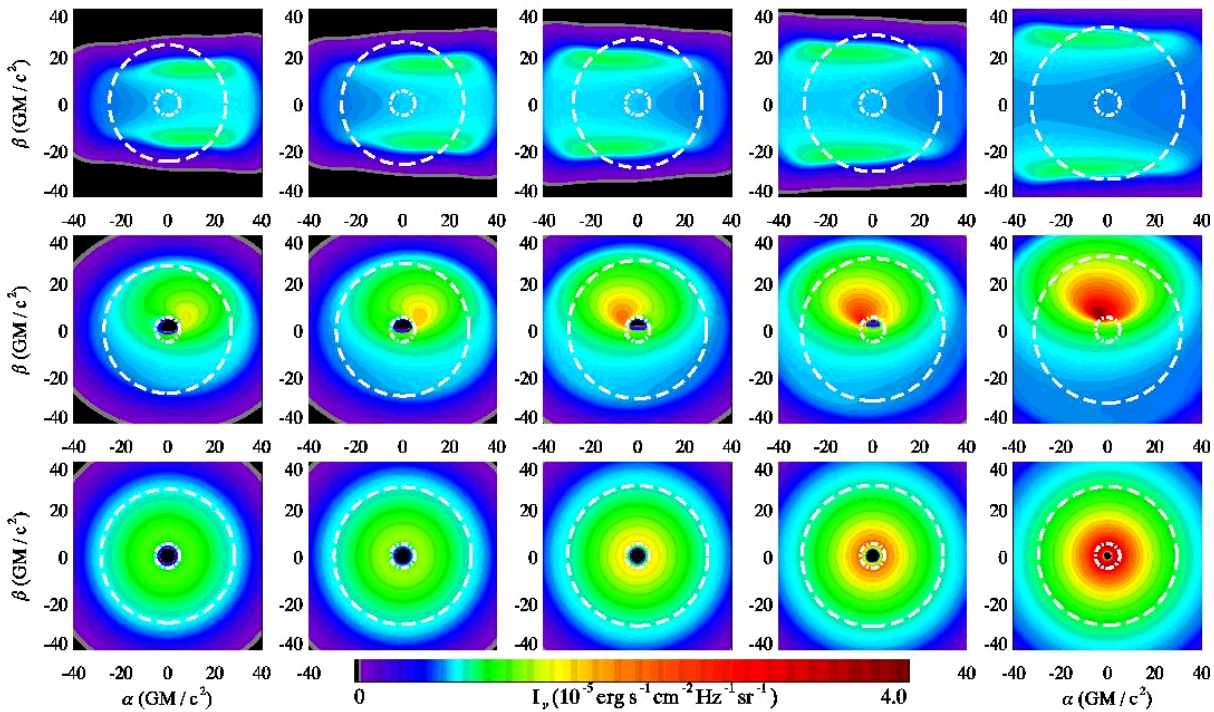}
\caption{Images at wavelength of 7.0mm of the ADAF of Kerr black holes
with spins $a=-0.998,-0.5.0,0.5,0.998$ (from left to right) and with the viewing
angles $\theta_{\rmscr{obs}}=90^{\rmscr{0}},45^{\rmscr{0}},0^{\rmscr{0}}$
(from top to bottom).
The spin of the black hole is along the perpendicular direction.
The diameter of the dash-dotted circle is twice the photon
capture impact parameter ($2 \sqrt{27}$M). The diameter of the dashed circle
is the intrinsic size ($D_{\rm ch}$) of each image, see definition in the text.
To show relative radiation intensity, the range of the intensity is scaled i
together for all the panels.
\label{fig:70}}
\end{figure}
\begin{figure}
\includegraphics[scale=.42]{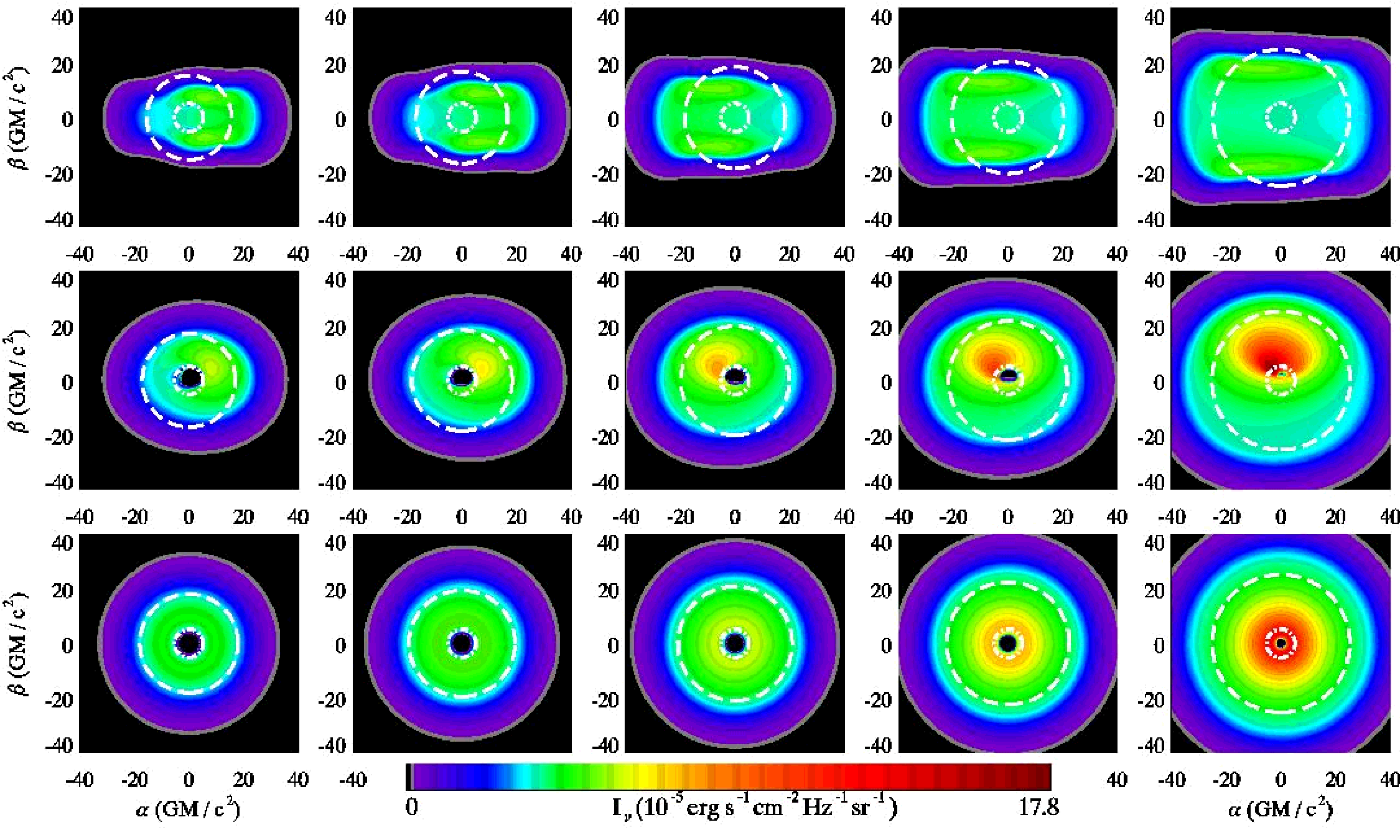}
\caption{As in Fig.\ref{fig:70}, but images at the wavelength of 3.5mm.
\label{fig:35}}
\end{figure}
\begin{figure}
\includegraphics[scale=.42]{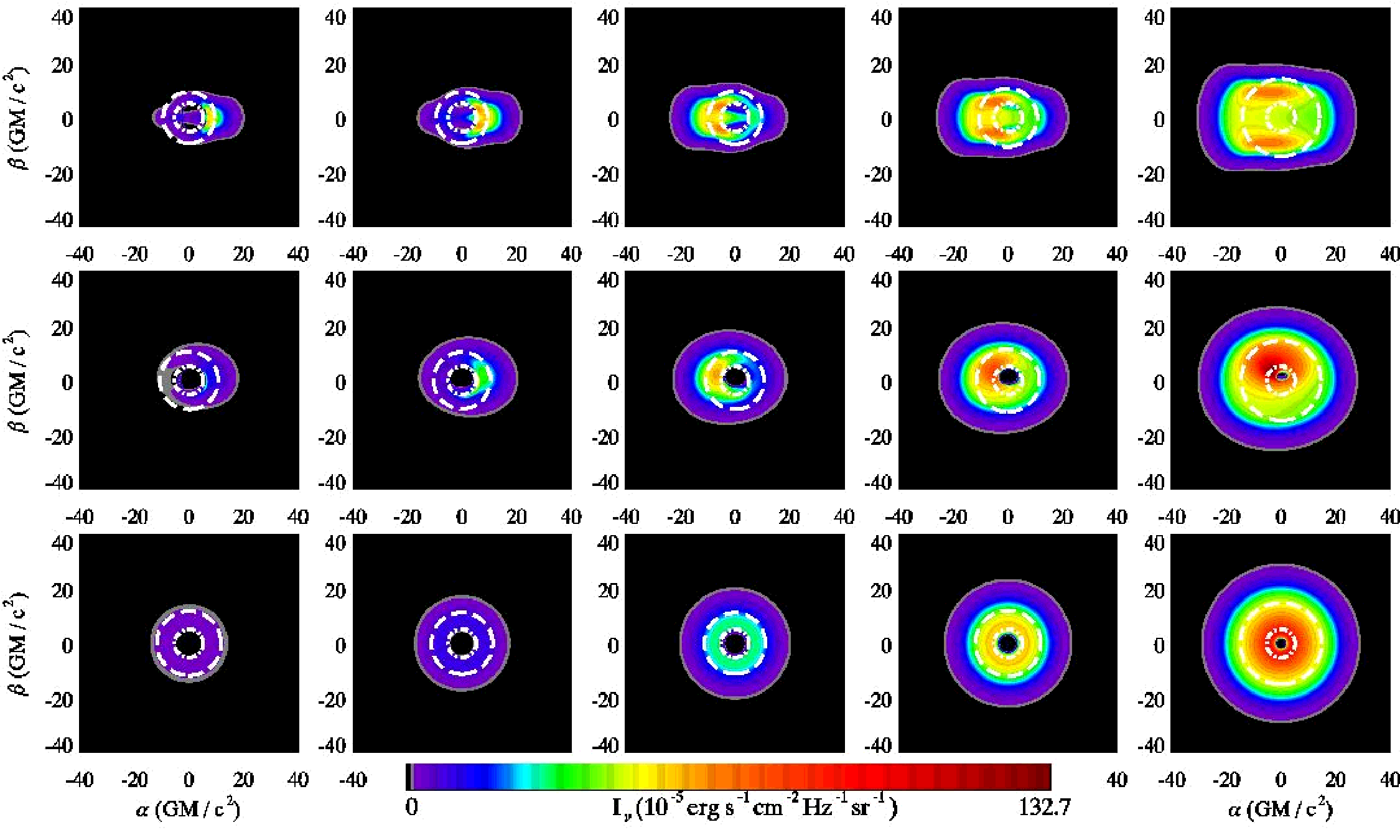}
\caption{As in Fig.\ref{fig:70}, but images at the wavelength of 1.3mm.  \label{fig:13}}
\end{figure}
Following the procedures described in the previous sections, we calculate
the images of GR-RIAF around a black hole with arbitrary spins,
inclination and at different observational wavelengths.
Our results are shown in Fig.\ref{fig_0}-\ref{fig:13}, respectively.

Figure \ref{fig_0} shows the results for face on case, i.e.,
the radiation intensity of GR-RIAF around a black hole as a function
of $\alpha$ (or $\beta$, because of the symmetry
of the system) for $a=0,\pm0.5,\pm0.998$ and for millimeter wavelengths,
$\lambda_{\rmscr{obs}}=7.0, 3.5, 1.3$ mm, in which we are most interested.
Here we define the intrinsic size $D_{\rm ch}$ of the disk
at a given frequency as
\begin{eqnarray}
    \int_{\rho \lesssim D_{\rm ch} /2} I_\nu(\alpha, \beta) d\Omega_{\rm obs} &=& \varepsilon \int I_\nu(\alpha, \beta) d\Omega_{\rm obs},
\end{eqnarray}
where $\rho = \sqrt{\alpha^2 + \beta^2}$ is the distance to 
the mass center of the hole,
$\Omega_{\rm obs}$ is the observed solid angle, and the coefficient is
the ratio of emission within the full width of half maximum (FWHM) of
a 2-dimensional(2-d) Gaussian function $G(\alpha, \beta)$ to its total
emission, i.e.,
\begin{eqnarray}
    \varepsilon &=& \frac{\int_{\rho \lesssim \rm FWHM /2} G(\alpha, \beta) d\Omega_{\rm obs}}{\int G(\alpha, \beta) d\Omega_{\rm obs}}.
\end{eqnarray}
By convention of {\it VLBI} observation, FWHM is used to measure the size
of a 2-d Gaussian distributed source.

For an image with the standard Gaussian distribution,
the intrinsic size $D_{\rm ch}$ defined here reduces to its
FWHM, therefore, for an image deviated from the Gaussian distribution,
the intrinsic size can be considered as its equivalent FWHM. 
As seen from Fig.\ref{fig_0},
the radiation intensity $I_{\nu}$ increases with the increase of BH
spin ranging from $a=-0.998$ to $a=0.998$, while the intrinsic size of the
disk, marked as region with radii covered by the horizontal lines for each case,
increases in the same way. These results are expected from the
effect of BH spin on the global solution of GR-RIAF (see
Fig.\ref{fig_adaf}): the larger the BH spin, the higher the
electron/ion number density and their temperature, and the larger
magnetic field in the disk. Because the emissivity at radio
wavebands is dominated by synchrotron radiation, larger
spin leads to higher radiation intensity of the disk. As has been
found in previous investigations, roughly the global solutions of
GR-RIAF for $a>0$ differ from that for $a=0$ more than the
corresponding $a<0$ cases \citep{1998ApJ...498..313G,1998ApJ...504..419P}
. This tendency is also reflected in
Fig.\ref{fig_0} for longer wavelength $\lambda_{\rmscr{obs}}=7.0,
3.5$mm. With the decrease of the wavelength, the emitting region
becomes smaller, because the physical conditions in the inner region
change dramatically, consequently, the difference among the final
intensity in the inner region becomes larger, especially for $a>0$
cases. The effect of the observational wavelength on the intrinsic
size of the images is also clearly shown in Fig.\ref{fig_0}: the
shorter the observational wavelength, the smaller the emitting
region of the disk.

For a Schwarzschild black hole, the maximal impact factor for photon
capture is $\sqrt{27}$M, therefore, the typical size of the BH
shadow is the circle with diameter of $2\sqrt{27}$M. The size of the
black hole shadow, region without any emission, are listed in
Table \ref{tb:shadow}. In the last column with $\lambda=1.3$mm, the
corresponding angular size is also listed.

In our calculation, the accretion flow extends to the horizon of the
hole $r_{\rm h^+}$. It is well known that the larger the spin of the
hole, the smaller its horizon and marginally stable orbit $r_{\rm ms}$.
This tendency is clearly shown in
Table \ref{tb:shadow}: 
it is found that the shadows of black hole decreases with the
increase of spin of the hole, if the disk is prograde with the hole,
whilst, if the disk is retrograde with the hole, the size of the shadow
does not change significantly.
For the same spin of the hole, the
size of the shadow at different wavelength does not change
significantly \citep[see also][]{2000ApJ...528L..13F,2006ApJ...636L.109B,2006MNRAS.367..905B,2004ApJ...611..996T}.

\begin{table}
\begin{center}
\caption{Size (diameter) of the shadow of the black hole\label{tb:shadow}}
\begin{tabular}{cccc}
\tableline\tableline
Spin of the hole & $\lambda=7.0mm$ & $\lambda=3.5mm$  & $\lambda=1.3mm$ \\
(a) & (M) & (M) & (M/$\mu$as) \\
\tableline
0.998 & 2.8 & 2.8 & 2.8/13.8 \\   
0.5 & 5.2 & 5.2 & 5.2/25.6  \\  
0 & 5.6 & 5.6 & 5.6/27.6 \\  
-0.5 & 6.0 & 6.0 & 6.4/31.5 \\ 
-0.998 & 6.4 & 6.8 & 7.6/37.5 \\  
\tableline
\end{tabular}
\end{center}
\end{table}

Figure \ref{fig:70}-\ref{fig:13} shows the images of the GR-RIAF
around a black hole with different spin $a$, the inclination angle
of the observer at infinity $\theta_{\rmscr{obs}}$, and at the
different wavelength $\lambda_{\rmscr{obs}}$. The observed images at
7.0mm for the prograde/retrograde disks with the black hole are
shown in Fig.\ref{fig:70}. Each column corresponds to the results for
different spin of the hole, from left to right, the spin of the hole
is $a=-0.998,-0.5,0,0.5,0.998$, respectively. Each rows corresponds to
the different $\theta_{\rmscr{obs}}$, from top to bottom,
$\theta_{\rmscr{obs}}=90^0,45^0,0^0$. For each image, its intrinsic
size is plotted in a white dashed circle, emission within which is
equal to $\varepsilon$ which is defined in Eq.~(29) times its 
total emission. The photon capture
region for a Schwarzschild black hole is also plotted in a white
dash-dotted circle for reference. All images in this figure are scaled in
the same way to show clearly the relative radiation intensity for
different spin and inclination.

In Figure \ref{fig:70}, for the same inclination, the
effects of spin on the size and intensity of the disk are similar:
the larger the BH spin, the larger the emitting region, i.e., the
intrinsic size, and the higher the radiation intensity. As
increasing the inclination angle, the shadow of the images decreases
due to the obscuration of the GR-RIAF with a finite height which is
about $H\leq r$. For the extreme case of
$\theta_{\rmscr{obs}}=90^0$, there is no shadow at all. When
$\theta_{\rmscr{obs}}=45^0$, the luminous part of the images which
is located at the left and top (for $a \ge 0$ cases), or right and
top (for $a<0$ cases) of the image is brighter than the luminous part
of the images for $\theta_{\rmscr{obs}}=0^0$. This is due to the
Lorentz boosting effects: at the left and top (or right and top) of
the disk, the fluid has the larger radial velocity to the observer
at infinity because of the rotation and radial motion of the
accretion flow. For $\theta_{\rmscr{obs}}=90^0$, the radio emission
is separated into two arcs for high spin (e.g. $a \ge
0.9$), while the weak radio emission appears between two arcs for
low spin. Our analysis is as follows: the photons from the
central part travel through the most dense region near the
equatorial plane of the disk, which results in the great
absorption of the emission, therefore, the resulting emission is
much weak. Radiative absorption is  the reason why the luminous
parts of the images
are not at the central regions near the equatorial plane of the
disk, and the intensity of the luminous parts is also lower than
that of the corresponding face on cases.
As discussed above, the intrinsic size
decreases with the increasing of the inclination angle. However,
such decrease becomes less significant for larger spin. For the
$a=0.998$ case, the intrinsic size nearly does not change. This can
be explained as follows: for larger inclination angle, the
obscuration of the disk decreases the emission, as well as the size
of the emitting region. On the other hand, the light bends more
significantly, bringing emission from the more inner region,
therefore both the emission and emitting region increase. For an
extremely-rotating ($a=0.998$) black hole, the emitting region
increased by its extremely curved space-time can almost compensate
that decreased by inclination, although the emission is not
compensated enough.

The images at 3.5mm and 1.3mm are shown in Fig.~\ref{fig:35} and
Fig.~\ref{fig:13}, respectively. The behavior of the images at
3.5mm and 1.3mm with varying the spin and the inclination are very
similar to those at 7.0mm.  The main differences are reflected in
the intrinsic sizes of images. At shorter wavelength, the disk
appears to be smaller than those at longer wavelength, this is
because the shorter wavelength photons are from the more inner
parts of the disk. For shorter wavelength photons emitting from
the inner parts of the disk, their optical depth might be so
significant, causing the shadow in the images of the disk even
when $\theta_{\rmscr{obs}}=90^0$ for low spin. The size of the
shadow is similar to the size of the circle with a radius of the
photon capture impact parameter $(\sqrt{27})$M. When the disk is
edge on, for the longer wavelength of the observed photons,
because the optical depth of the emitted photon near the
equatorial plane is not so large that the shadow does not appear.
For $\lambda_{\rmscr{obs}}=1.3$mm and median spin $a \leq 0.5$,
the luminous parts of the images are brighter than these of the
corresponding face on cases, so that the larger inclination results in
higher emission. This is because the Lorentz boosting effect
dominates the radiative absorption. Exceptionally, at high
spin, such as $a=0.998$ and $\theta_{\rmscr{obs}}=90^0$, the
emitting region is near to the horizon of the black hole, thus, the
radiative absorption dominates the Lorentz boosting, as a result,
the intensity of the luminous parts is lower than that of the
corresponding face on cases as in  Fig.~\ref{fig:70} and
Fig.~\ref{fig:35} for longer wavelengths.

\section{Discussions}

In this work, the images of the GR-RIAF around a Kerr black hole
are simulated. The main results are summarized as follows: 1) the
effects of BH spin: the larger the spin, the smaller the size of
black hole shadow, which is similar to the previous results
\citep[e.g.,][]{2000ApJ...528L..13F,2006ApJ...636L.109B,2006MNRAS.367..905B,2004ApJ...611..996T}).
The larger the spin, the larger the intrinsic size of the image,
and the brighter the whole disk, especially its inner part. These
effects are more significant in a prograde disk than a retrograde
disk. We should emphasize that this contrasts with the result
obtained in \citep{2006ApJ...636L.109B,2006MNRAS.367..905B}. The
reason is that \citet{2006ApJ...636L.109B,2006MNRAS.367..905B}
assumed a parameterized ADAF solution for a spinning black hole
which is required to fit the observed spectrum, therefore, they
found that the thermal electron density must decrease with
increasing spin, as well with increasing inclination.  More
important is that the non-thermal electron density must increase,
in order to produce the radio spectrum. However, in this work, we
do all the calculation in the light of a fiducial GR-ADAF model in
which the accretion rate is fixed. Consequently, the electron
density increase with increasing spin (see Fig.~\ref{fig_adaf}). 
2) the effects of the inclination: the larger the
inclination angles, the smaller the size of black hole shadow, and
the smaller the intrinsic size of the image. This is
consistent with \citep{2007CQGra..24..259N}. If the disk is edge
on, the shadow are even obscured at all. The central part of the
image is both darkened by the absorption and brightened by the
light bending, so that how the brightness of the image changes
depends on the observational wavelength. 3) the effects of the
observational wavelength: the shorter the observational
wavelength, the smaller the intrinsic size of the image.

Although a larger spin makes the intrinsic size enlarged, it makes the
total emission enhanced to a greater extent.
This tendency differs from face-on to edge-on cases.
At 7.0mm, for
instance, as shown in Fig.\ref{fig:70}, an extremely-rotating
black hole ($a=0.998$) enhances $> 60\%$ total emission than a
non-rotating black hole ($a=0$) does. For the face-on case, however, the
intrinsic size is enlarged only by $\approx 15\%$. For the same
comparison at 3.5mm, as shown in Fig.\ref{fig:35}, the total emission
is enhanced by $\approx 99\%$ while the intrinsic size is enlarged
by only $\approx 23\%$. Therefore, more emission comes from the
inner part of the image if the black hole spin increases.

As discussed in \S 5, in \citet{2006ApJ...642L..45Y}, the
global structure of the RIAF were calculated by adopting
Pseudo-Newtonian potential, and taking the non-thermal electrons
into consideration. The disk is viewed as facing on, and the
thickness of the disk is ignored. It was found that the scattered
images are more extended than the observational images at both
7.0mm and 3.5mm \citep{2005Natur.438...62S}. 
In our GR-RIAF model with the same parameters,
the scattered images are even bigger. This is because the lower
electron temperature and density are obtained in GR-RIAF model,
enlarging the emission region.  However, the overestimated
intrinsic sizes do not imply an anti-rotating black hole, which
predicts smaller intrinsic sizes.  As mentioned above, this is 
because the anti-rotating black hole also predicts much lower 
total emission. 
The required emitting region should be enlarged a lot to
produce the observed emission. On the other hand, a co-rotating
black hole is preferred, since the required emitting region could
be more compact.

\begin{figure}
\includegraphics[scale=.98]{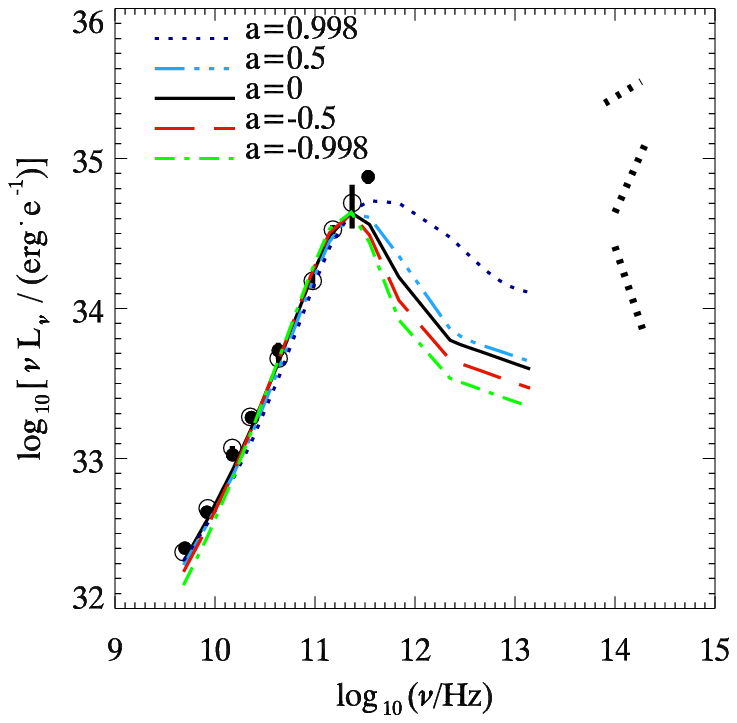}
\includegraphics[scale=.45]{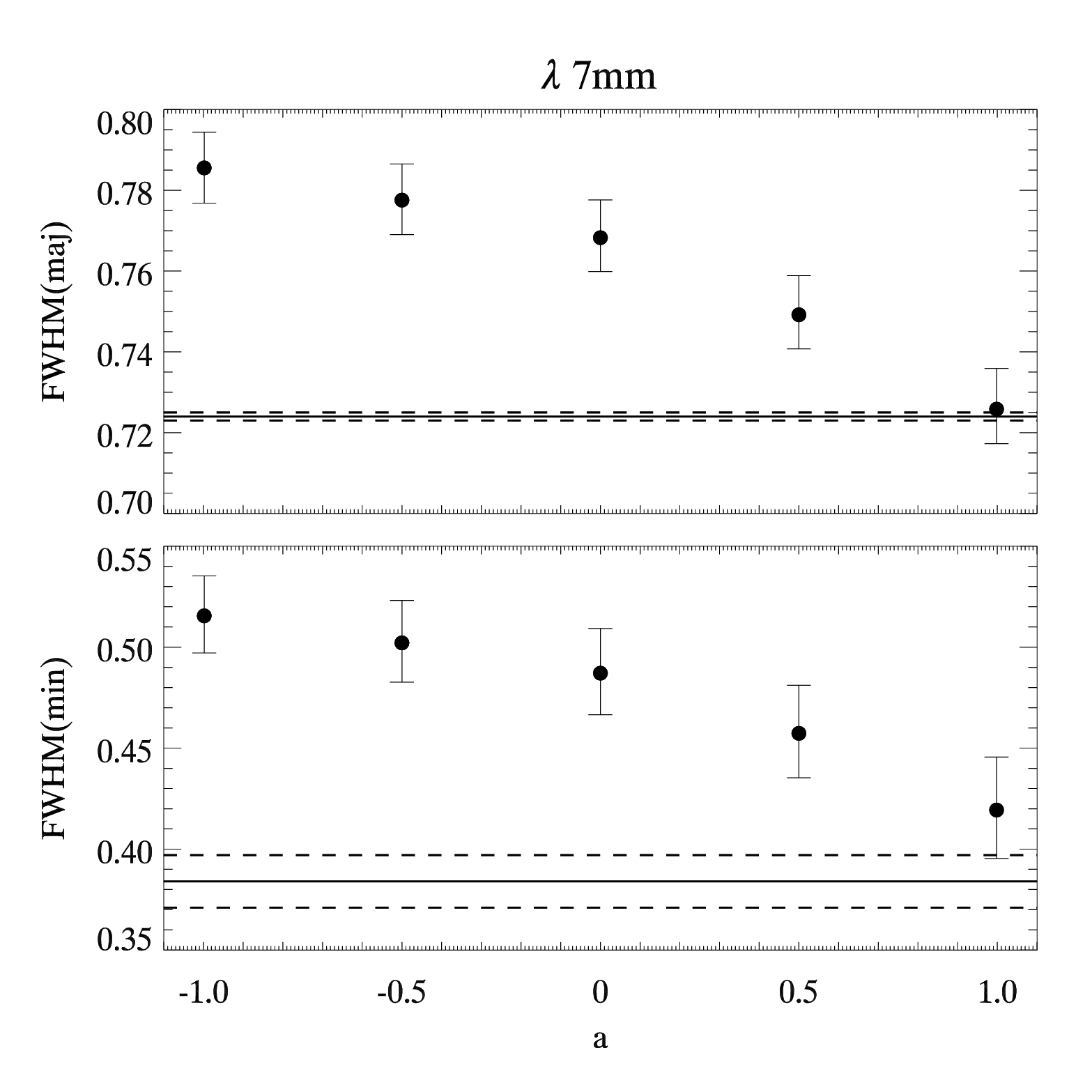} \\
\includegraphics[scale=.98]{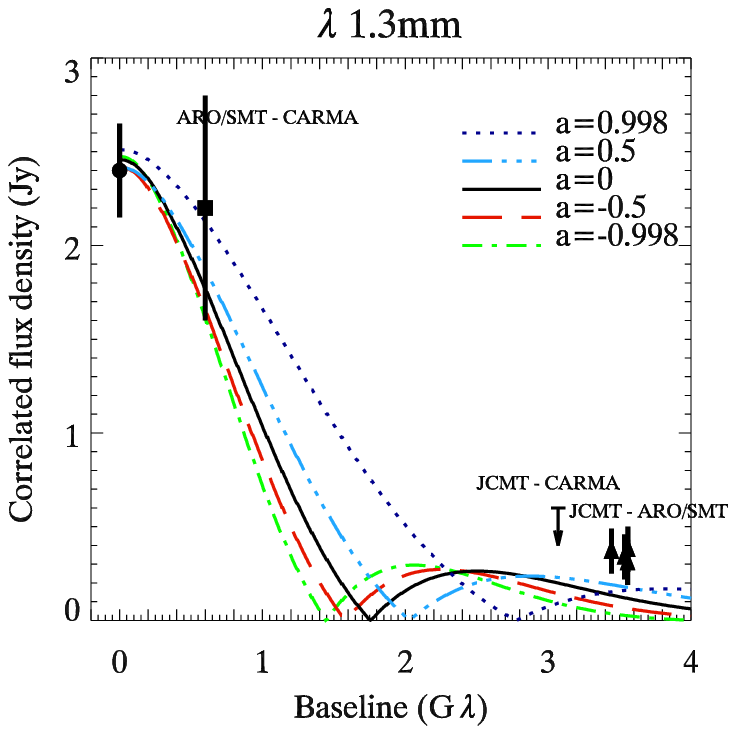}
\includegraphics[scale=.45]{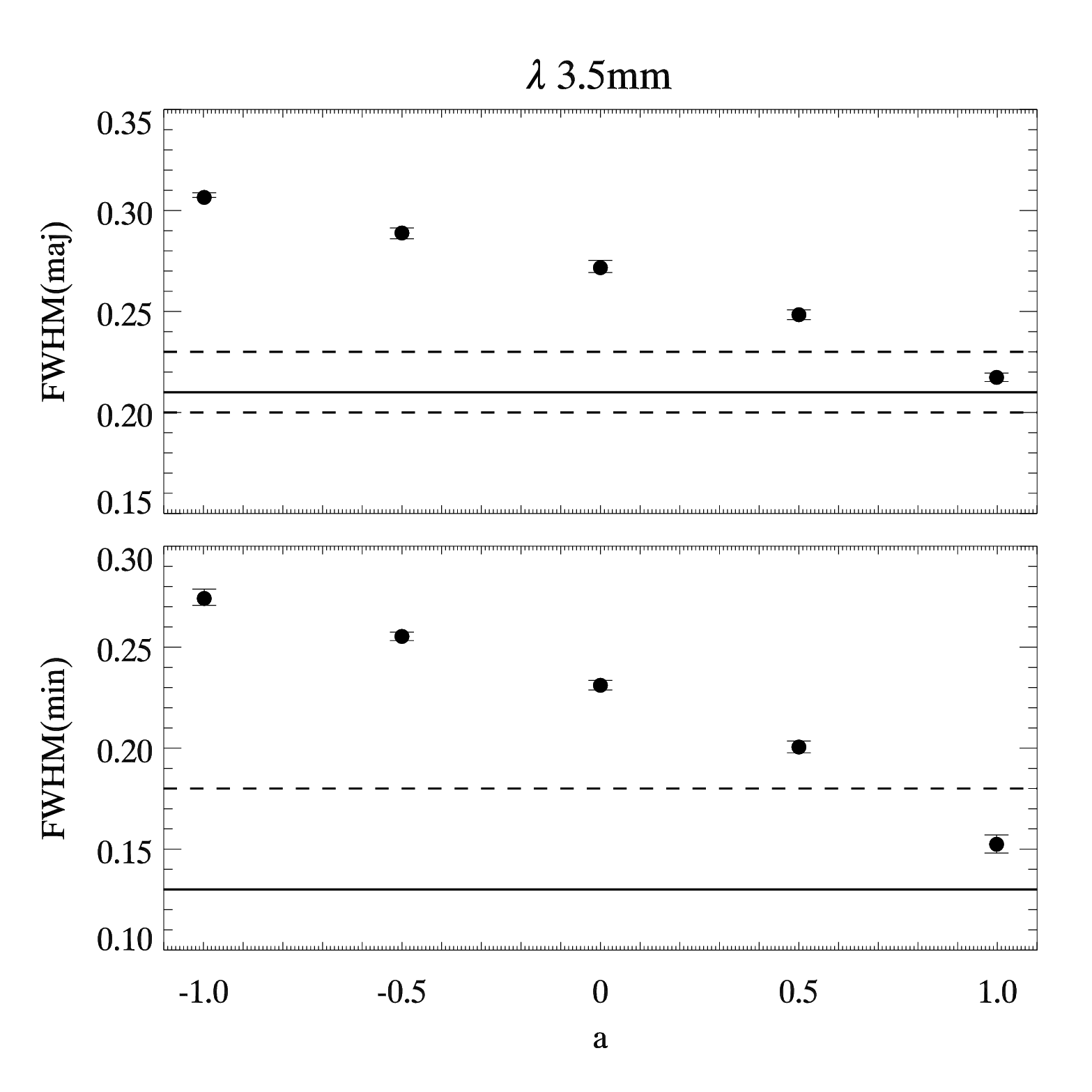}
\caption{ Spectra, full-width of half maximum of the
scatter-broadened image at 7.0mm and 3.5mm, and visibility
profiles at 1.3mm of Sgr A*, predicted by face-on GR-RIAF with
$a=-0.998,0.5,0,0.5$ and $0.998$. See the details in the text.
\label{appl}}
\end{figure}
For a primary application, we adjust the index of the wind
index $s$ and energy fraction of non-thermal electrons
$\eta$ to reproduce the spectrum of quiescent state of Sgr A*.
Only face-on case is investigated as an example. For the
Schwarzschild case, we choose $s=0.285$ and $\eta=0.4\%$,
different from those in \citet{2006ApJ...642L..45Y} due to the
relativistic effects. In order to produce the observed
the spectrum both in radio and sub-millimeter band,
$s$ should be decreased, while $\eta$ should be increased for a
faster co-rotating black hole. Contrarily, for a faster
anti-rotating black hole, $s$ should be increased
while $\eta$ should be decreased for the same purpose. 
In Fig.\ref{appl}, the predicted spectra, the FWHM 
of the images at
7.0mm and 3.5mm, and the visibility profiles at 1.3mm are
shown for five examples with different spins, with the
interstellar scattering effect having been taken into consideration.
As expected, the smaller sizes are
obtained at all the three wavelengths for a higher spin.
Compared with the measured sizes at 7.0mm and 3.5mm by
\citet{2005Natur.438...62S},
the predicted FWHMs are plotted in the solid and dashed lines for
averages and limits in the right two panels of Fig.\ref{appl},
respectively. It is strongly indicated in Fig.\ref{appl}
that a fast co-rotating black hole is preferred, 
if the disk is facing on. Notice that the FWHM of the
minor axis at 7.0mm overestimates the observation, even with an
extremely rotating black hole. This may
imply an inclined disk which makes the intrinsic sizes smaller
(see section 5). By analyzing the existing spectral and polariztion
measurements, \citet{2008arXiv0809.4490B} found that 
low values of the inclination angle are excluded,
which is consistent with our analysis.

The {\it VLBI} observations also predict that the black hole
shadow structure is hoped to be resolved if the observational
wavelength decreases to 1.3mm. The scattering size extrapolated
from previous longer wavelength observations is $\sim 23\mu$as,
which means any structure with size larger than it would not be
washed out. As shown in the bottom-left panel of
Fig.\ref{appl}, the long baseline is predicted for the
existence of the null point,
which would be an evidence of the shadow region of a black hole with
high spin. The high spin causes the emission region more compact.
Compared to the latest visibility detections at 1.3mm reported by
\citet{2008Natur.455...78D}, only two of the five cases, with
$a=0.5$ and $0.998$, marginally reproduce the correlated density
at $\sim 3.5 {\rm G}\lambda$ detected by the $JCMT-ARO/SMT$ baseline.
These results also suggest that the black hole is fast rotating or the
inclination angle is large, which is consistent with what suggested
by the previous observations at longer wavelengths.  As shown in table
\ref{tb:shadow}, a fast rotating black hole with $a>0.5$ predicts
size of shadow less than the scattering size, so that its shadow
structure is hard to resolved,
especially for a highly-inclined disk. If the Galactic
central black hole is really fast rotating, the future shorter wavelengths
(sub-millimeter) are necessary for the detections.

\acknowledgments We would like to thank Dr. Avery E. Broderick for
useful discussion. This work is partially supported by National Basic
Research Program of China (2009CB824800), the National Natural
Science Foundation (10733010, 10773020, 10673010, 10573016,
10833002, 10821302), the CAS (grant KJCX2-YW-T03), and Program for
New Century Excellent Talents in University.


\begin{thebibliography}{39}
\expandafter\ifx\csname natexlab\endcsname\relax\def\natexlab#1{#1}\fi

\bibitem[{{Abramowicz} {et~al.}(1997){Abramowicz}, {Lanza}, \&
  {Percival}}]{1997ApJ...479..179A}
{Abramowicz}, M.~A., {Lanza}, A., \& {Percival}, M.~J. 1997, \apj, 479, 179

\bibitem[{{Bardeen} {et~al.}(1972){Bardeen}, {Press}, \&
  {Teukolsky}}]{1972ApJ...178..347B}
{Bardeen}, J.~M., {Press}, W.~H., \& {Teukolsky}, S.~A. 1972, \apj, 178, 347

\bibitem[{{Blandford} \& {Begelman}(1999)}]{1999MNRAS.303L...1B}
{Blandford}, R.~D. \& {Begelman}, M.~C. 1999, \mnras, 303, L1

\bibitem[{{Broderick} {et~al.}(2008){Broderick}, {Fish}, {Doeleman}, \&
  {Loeb}}]{2008arXiv0809.4490B}
{Broderick}, A.~E., {Fish}, V.~L., {Doeleman}, S.~S., \& {Loeb}, A. 2008,
  ArXiv:0809.4490, ApJ in press

\bibitem[{{Broderick} \& {Loeb}(2005)}]{2005MNRAS.363..353B}
{Broderick}, A.~E. \& {Loeb}, A. 2005, \mnras, 363, 353

\bibitem[{{Broderick} \& {Loeb}(2006{\natexlab{a}})}]{2006ApJ...636L.109B}
---. 2006{\natexlab{a}}, \apjl, 636, L109

\bibitem[{{Broderick} \& {Loeb}(2006{\natexlab{b}})}]{2006MNRAS.367..905B}
---. 2006{\natexlab{b}}, \mnras, 367, 905

\bibitem[{{Cadez} {et~al.}(1998){Cadez}, {Fanton}, \&
  {Calvani}}]{1998NewA....3..647C}
{Cadez}, A., {Fanton}, C., \& {Calvani}, M. 1998, New Astronomy, 3, 647

\bibitem[{{Carter}(1968)}]{1968PhRv..174.1559C}
{Carter}, B. 1968, Physical Review, 174, 1559

\bibitem[{{Chandrasekhar}(1983)}]{1983mtbh.book.....C}
{Chandrasekhar}, S. 1983, {The mathematical theory of black holes} (
  Oxford: Clarendon Press)

\bibitem[{{Cunningham} \& {Bardeen}(1973)}]{1973ApJ...183..237C}
{Cunningham}, J.~M. \& {Bardeen}, C.~T. 1973, \apj, 183, 237

\bibitem[{{Doeleman} \& {Bower}(2004)}]{2004GCNew..18....6D}
{Doeleman}, S. \& {Bower}, G. 2004, Galactic Center Newsletter, 18, 6

\bibitem[{{Doeleman} {et~al.}(2008){Doeleman}, {Weintroub}, {Rogers},
  {Plambeck}, {Freund}, {Tilanus}, {Friberg}, {Ziurys}, {Moran}, {Corey},
  {Young}, {Smythe}, {Titus}, {Marrone}, {Cappallo}, {Bock}, {Bower},
  {Chamberlin}, {Davis}, {Krichbaum}, {Lamb}, {Maness}, {Niell}, {Roy},
  {Strittmatter}, {Werthimer}, {Whitney}, \& {Woody}}]{2008Natur.455...78D}
{Doeleman}, S.~S., {Weintroub}, J., {Rogers}, A.~E.~E., \etal
  2008, \nat, 455, 78

\bibitem[{{Ferrarese} \& {Merritt}(2000)}]{2000ApJ...539L...9F}
{Ferrarese}, L. \& {Merritt}, D. 2000, \apjl, 539, L9

\bibitem[{{Fish} {et~al.}(2008)}] {2008arXiv0809.4489F}
{Fish}, V.~L., {Broderick}, A.~E., {Doeleman}, S.~S., \& {Loeb}, A. 2008,
  ArXiv:0809.4489

\bibitem[{{Fuerst} \& {Wu}(2004)}]{2004A&A...424..733F}
{Fuerst}, S.~V. \& {Wu}, K. 2004, \aap, 424, 733

\bibitem[{{Gammie} \& {Popham}(1998)}]{1998ApJ...498..313G}
{Gammie}, C.~F. \& {Popham}, R. 1998, \apj, 498, 313

\bibitem[{{Gebhardt} {et~al.}(2000)}]{2000ApJ...539L..13G}
{Gebhardt}, K., {Bender}, R., {Bower}, G., \etal
  2000, \apjl, 539, L13

\bibitem[{{Ghez} {et~al.}(2005){Ghez}, {Salim}, {Hornstein}, {Tanner}, {Lu},
  {Morris}, {Becklin}, \& {Duch{\^e}ne}}]{2005ApJ...620..744G}
{Ghez}, A.~M., {Salim}, S., {Hornstein}, S.~D., \etal
  2005, \apj, 620, 744

\bibitem[Ghez et al.(2008)]{2008ApJ...689.1044G} Ghez, A.~M., et al.\ 2008,
\apj, 689, 1044

\bibitem[Falcke et al.(2000)]{2000ApJ...528L..13F} Falcke, H., Melia, F.,
\& Agol, E.\ 2000, \apjl, 528, L13

\bibitem[{{Hopkins} {et~al.}(2008)}]{2008ApJS..175..356H}
{Hopkins}, P.~F., {Hernquist}, L., {Cox}, T.~J., \& {Kere{\v s}}, D. 2008,
  \apjs, 175, 356

\bibitem[{{Huang} {et~al.}(2007){Huang}, {Cai}, {Shen}, \&
  {Yuan}}]{2007MNRAS.379..833H}
{Huang}, L., {Cai}, M., {Shen}, Z.-Q., \& {Yuan}, F. 2007, \mnras, 379, 833

\bibitem[{{Huang} {et~al.}(2008){Huang}, {Liu}, {Shen}, {Cai} \etal
}]{2008ApJ...676L.119H}
{Huang}, L., {Liu}, S., {Shen}, Z.-Q., {Cai}, M.~J. \etal
2008, \apjl, 676, L119

\bibitem[{{Kormendy} \& {Richstone}(1995)}]{1995ARA&A..33..581K}
{Kormendy}, J. \& {Richstone}, D. 1995, \araa, 33, 581

\bibitem[{{Manmoto}(2000)}]{2000ApJ...534..734M}
{Manmoto}, T. 2000, \apj, 534, 734

\bibitem[{{Maoz}(1998)}]{1998ApJ...494L.181M}
{Maoz}, E. 1998, \apjl, 494, L181

\bibitem[{{Miller}(2006)}]{2006MNRAS.367L..32M}
{Miller}, M.~C. 2006, \mnras, 367, L32

\bibitem[{{Miyoshi} {et~al.}(2004){Miyoshi}, {Ishitsuka}, {Kameno} \etal
  }]{2004PThPS.155..186M}
{Miyoshi}, M., {Ishitsuka}, J.~K., {Kameno}, S. \etal
  2004, Progress of Theoretical Physics Supplement, 155, 186

\bibitem[{{Munyaneza} {et~al.}(1998){Munyaneza}, {Tsiklauri}, \&
  {Viollier}}]{1998ApJ...509L.105M}
{Munyaneza}, F., {Tsiklauri}, D., \& {Viollier}, R.~D. 1998, \apjl, 509, L105

\bibitem[{{Narayan}(2002)}]{2002luml.conf..405N}
{Narayan}, R. 2002, in Lighthouses of the Universe: The Most Luminous Celestial
  Objects and Their Use for Cosmology, ed. M.~{Gilfanov}, R.~{Sunyeav}, \&
  E.~{Churazov} (Berlin: Springer)

\bibitem[{{Narayan} {et~al.}(1998){Narayan}, {Mahadevan}, {Grindlay}
 \etal}]{1998ApJ...492..554N}
{Narayan}, R., {Mahadevan}, R., {Grindlay}, J.~E. \etal  1998, \apj, 492, 554

\bibitem[{{Narayan} {et~al.}(1995){Narayan}, {Yi}, \&
  {Mahadevan}}]{1995Natur.374..623N}
{Narayan}, R., {Yi}, I., \& {Mahadevan}, R. 1995, \nat, 374, 623

\bibitem[Noble et al.(2007)]{2007CQGra..24..259N} Noble, S.~C., Leung,
P.~K., Gammie, C.~F.,
\& Book, L.~G.\ 2007, Classical and Quantum Gravity, 24, 259

\bibitem[{{Paumard} {et~al.}(2005){Paumard}, {Perrin}, {Eckart}, {Genzel},
  {Lena}, {Schoedel}, {Eisenhauer}, {Mueller}, \&
  {Gillessen}}]{2005AN....326..568P}
{Paumard}, T., {Perrin}, G., {Eckart}, A., \etal
  2005, Astronomische Nachrichten, 326, 568

\bibitem[{{Peitz} \& {Appl}(1997)}]{1997MNRAS.286..681P}
{Peitz}, J. \& {Appl}, S. 1997, \mnras, 286, 681

\bibitem[{{Popham} \& {Gammie}(1998)}]{1998ApJ...504..419P}
{Popham}, R. \& {Gammie}, C.~F. 1998, \apj, 504, 419

\bibitem[{{Rauch} \& {Blandford}(1994)}]{1994ApJ...421...46R}
{Rauch}, K.~P. \& {Blandford}, R.~D. 1994, \apj, 421, 46

\bibitem[{{Rybicki} \& {Lightman}(1979)}]{1979rpa..book.....R}
{Rybicki}, G.~B. \& {Lightman}, A.~P. 1979, {Radiative processes in
  astrophysics} (New York, Wiley-Interscience)

\bibitem[{{Sch{\"o}del} {et~al.}(2003){Sch{\"o}del}, {Ott}, {Genzel}, {Eckart},
  {Mouawad}, \& {Alexander}}]{2003ApJ...596.1015S}
{Sch{\"o}del}, R., {Ott}, T., {Genzel}, R., \etal 2003, \apj, 596, 1015

\bibitem[{{Shen} {et~al.}(2005){Shen}, {Lo}, {Liang} \etal }]
{2005Natur.438...62S} {Shen}, Z.-Q., {Lo}, K.~Y., {Liang}, M.-C. \etal
  2005, \nat, 438, 62

\bibitem[Takahashi(2004)]{2004ApJ...611..996T} Takahashi, R.\ 2004, \apj,
611, 996

\bibitem[{{Torres} {et~al.}(2000){Torres}, {Capozziello}, \&
  {Lambiase}}]{2000PhRvD..62j4012T}
{Torres}, D.~F., {Capozziello}, S., \& {Lambiase}, G. 2000, \prd, 62, 104012

\bibitem[{{Tsiklauri} \& {Viollier}(1998)}]{1998ApJ...500..591T}
{Tsiklauri}, D. \& {Viollier}, R.~D. 1998, \apj, 500, 591

\bibitem[{{Yuan} {et~al.}(2003){Yuan}, {Quataert}, \&
  {Narayan}}]{2003ApJ...598..301Y}
{Yuan}, F., {Quataert}, E., \& {Narayan}, R. 2003, \apj, 598, 301

\bibitem[{{Yuan} {et~al.}(2004){Yuan}, {Quataert}, \&
  {Narayan}}]{2004ApJ...606..894Y}
---. 2004, \apj, 606, 894

\bibitem[{{Yuan}, {Narayan}, \& {Rees}(2004){Yuan}, {Narayan}, \&
  {Rees}}]{2004ApJ...606.1112Y}
{Yuan}, Y.-F., {Narayan}, R., \& {Rees}, M.~J. 2004, \apj, 606, 1112

\bibitem[{{Yuan} {et~al.}(2006){Yuan}, {Shen}, \&
  {Huang}}]{2006ApJ...642L..45Y}
{Yuan}, F., {Shen}, Z.-Q., \& {Huang}, L. 2006, \apjl, 642, L45



\end{thebibliography}

\clearpage
\appendix
\section{The reference frames}
In the ADAF model, the accretion flow has both azimuthal and radial motion.
Following Peitz \& Appl (1997) (see also Gammie \& Popham 1998, Manmoto etal
2000), we introduce three reference frames to show how to describe the
velocity fields of accreted gas. The first is the locally non-rotating frame
(LNRF) which is also called zero angular momentum observer (ZAMO), an
orthonormal tetrad basis carried by observers whose four velocity in
Boyer-Lindquist coordinates is $u^{\mu}=e^{-\nu}(1,0,0,\omega)$.
The four contra-variant basis
vectors of LNRF compose a matrix as follows,
\bea
e^{\mu}_{\, .\,\, (a)}(\rmmat{LNRF})  & = &
 \left[
\begin{array}{cccc}
e^{-\nu} & 0 & 0 & e^{-\nu} \omega\\
0 & e^{-\mu_1} & 0 & 0 \\
0 & 0 & e^{-\mu_2} & 0  \\
0 & 0 & 0 & e^{-\psi} \\
\end{array}     \right] ,
\eea
while the four covariant basis vectors compose a matrix as follows,
\bea
e_{\mu}^{\, .\,\, (a)}(\rmmat{LNRF})  & = &
 \left[
\begin{array}{cccc}
e^{\nu} & 0 & 0 & 0 \\
0 & e^{\mu_1} & 0 & 0 \\
0 & 0 & e^{\mu_2} & 0  \\
-\omega e^{\psi} & 0 & 0 & e^{\psi} \\
\end{array}     \right] .
\eea
As well known, the four rows of the matrix give four vectors in
the curved space-time, whilst the four columns give four vectors
in the local inertia frame, i.e. four vectors in the 'Minkovski'
space-time.

The second frame is the so-call co-rotating frame, another
orthonormal tetrad basis carried by observers co-rotates with the
fluid with the coordinate angular velocity $\Omega=u^{\phi}/u^t$,
i.e. whose four velocity is $u^{\mu}=u^t(1,0,0,\Omega)$. The
relative physical azimuthal velocity of CRF with respect to LNRF
is $V^{(\phi)}=e^{\phi-\nu}(\Omega-\omega)$, therefore, the four
basis vectors of CRF can be obtained by an azimuthal Lorentz boost
from these of LNRF, \bea
e^{\mu}_{\, .\,\, (a)}(\rmmat{CRF})  &=& \Lambda_a^{\, .\,\,b} \,\, e^{\mu}_{\, .\,\, (b)}(\rmmat{LNRF}) \nonumber \\
&=&
\left[
\begin{array}{cccc}
\gamma_{\phi} & 0 & 0 & \gamma_{\phi} \beta_{\phi}\\
0 & 1 & 0 & 0 \\
0 & 0 & 1 & 0  \\
\gamma_{\phi} \beta_{\phi} & 0 & 0 &  \gamma_{\phi} \\
\end{array} \right]
 \left[
\begin{array}{cccc}
e^{-\nu} & 0 & 0 & e^{-\nu} \omega\\
0 & e^{-\mu_1} & 0 & 0 \\
0 & 0 & e^{-\mu_2} & 0  \\
0 & 0 & 0 & e^{-\psi} \\
\end{array}     \right] , \nonumber \\
&=&
 \left[
\begin{array}{cccc}
\gamma_{\phi}e^{-\nu}& 0 & 0 & \gamma_{\phi} (\omega e^{-\nu}+\beta_{\phi} e^{-\psi})\\
0 & e^{-\mu_1} & 0 & 0 \\
0 & 0 & e^{-\mu_2} & 0  \\
\gamma_{\phi} \beta_{\phi}e^{-\nu} & 0 & 0 &  \gamma_{\phi} (\beta_{\phi}\omega e^{-\nu} + e^{-\psi})\\
\end{array} \right].
\\
e_{\mu}^{\, .\,\, (a)}(\rmmat{CRF})  &=& \Lambda^a_{\, .\,\,b} \,\, e_{\mu}^{\, .\,\, (b)}(\rmmat{LNRF}) \nonumber \\
&=&
 \left[
\begin{array}{cccc}
\gamma_{\phi} (e^{\nu}+\beta_{\phi} \omega e^{\psi})& 0 & 0 & -\gamma_{\phi} \beta_{\phi}e^{\psi}\\
0 & e^{\mu_1} & 0 & 0 \\
0 & 0 & e^{\mu_2} & 0  \\
-\gamma_{\phi} (\beta_{\phi}e^{\nu}+\omega e^{\psi})& 0 & 0 &  \gamma_{\phi} e^{\psi}\\
\end{array} \right].
\eea

The final frame is the local rest frame (LRF) of the fluid,
another orthonormal tetrad basis carried by observers moving
with the fluid,
i.e. whose four velocity is $u^{\mu}=(u^t,u^r,0,u^{\phi})$.
Suppose the relative physical radial velocity of LRF with respect to
CRF is $V^{(r)}\equiv V$, therefore, the four
basis vectors of LRF can be obtained by a radial Lorentz boost from
these of CRF,
\bea
e^{\mu}_{\, .\,\, (a)}(\rmmat{LRF})  &=& \Lambda_a^{\, .\,\,b} \,\, e^{\mu}_{\, .\,\, (b)}(\rmmat{CRF}) \nonumber \\
&=&
\left[
\begin{array}{cccc}
\gamma_r & \gamma_r \beta_r & 0 & 0\\
\gamma_r \beta_r & \gamma_r & 0 & 0 \\
0 & 0 & 1 & 0  \\
0 & 0 & 0 & 1  \\
\end{array} \right]
 \left[
\begin{array}{cccc}
\gamma_{\phi}e^{-\nu}& 0 & 0 & \gamma_{\phi} (\omega e^{-\nu}+\beta_{\phi} e^{-\psi})\\
0 & e^{-\mu_1} & 0 & 0 \\
0 & 0 & e^{-\mu_2} & 0  \\
\gamma_{\phi} \beta_{\phi}e^{-\nu} & 0 & 0 &  \gamma_{\phi} (\beta_{\phi}\omega e^{-\nu} + e^{-\psi})\\
\end{array} \right]. \nonumber \\
&=&
 \left[
\begin{array}{cccc}
\gamma_r\gamma_{\phi}e^{-\nu}& \gamma_r \beta_r e^{-\mu_1}& 0 &
\gamma_r \gamma_{\phi} (\omega e^{-\nu}+\beta_{\phi} e^{-\psi})\\
\gamma_r\gamma_{\phi} \beta_r e^{-\nu} & \gamma_r e^{-\mu_1}& 0 &
\gamma_r \gamma_{\phi} \beta_r(\omega e^{-\nu}+\beta_{\phi} e^{-\psi}) \\
0 & 0 & e^{-\mu_2} & 0  \\
\gamma_{\phi} \beta_{\phi}e^{-\nu} & 0 & 0 &  \gamma_{\phi} (\beta_{\phi}\omega e^{-\nu} + e^{-\psi})\\
\end{array} \right].
\\
e_{\mu}^{\, .\,\, (a)}(\rmmat{LRF})  &=& \Lambda^a_{\, .\,\,b} \,\, e_{\mu}^{\, .\,\, (b)}(\rmmat{CRF}) \nonumber \\
&=&
 \left[
\begin{array}{cccc}
\gamma_r \gamma_{\phi} (e^{\nu}+\beta_{\phi} \omega e^{\psi})& -\gamma_r \beta_r
e^{\mu_1} & 0 & -\gamma_r \gamma_{\phi} \beta_{\phi}e^{\psi}\\
-\gamma_r \gamma_{\phi} \beta_r(e^{\nu}+\beta_{\phi} \omega e^{\psi})
& \gamma_re^{\mu_1} & 0 & \gamma_r \gamma_{\phi} \beta_r \beta_{\phi} e^{\psi}\\
0 & 0 & e^{\mu_2} & 0  \\
-\gamma_{\phi} (\beta_{\phi}e^{\nu}+\omega e^{\psi})& 0 & 0 &  \gamma_{\phi} e^{\psi}\\
\end{array} \right].
\eea

\end{document}